\documentclass[10pt,aps,prb,twocolumn,preprintnumbers,amsmath,amssymb,floatfix,showpacs,citeautoscript,superscriptaddress]{revtex4-2}
\usepackage[utf8]{inputenc}
\usepackage[T1]{fontenc}
\usepackage[english]{babel}
\usepackage[pdftex]{graphicx}
\usepackage[normalem]{ulem}
\usepackage{multirow}
\usepackage{amsmath}

\usepackage{times}
\usepackage{color}
\usepackage[usenames,dvipsnames,svgnames,table]{xcolor}
\usepackage[colorlinks,plainpages=false,linkcolor=blue,urlcolor=blue,citecolor=blue,pdfpagemode=UseNone]{hyperref}
\usepackage{svg}
\usepackage{breqn}

\renewcommand{\AA}{\text{\r{A}}}

\newcommand{\uf}{\mathrm{UF}^3}

\definecolor{d12orange}{rgb}{0.8500    0.3250    0.0980}
\definecolor{g8yellow}{rgb}{0.    0.6    0.298}
\definecolor{ppurple}{rgb}{0.4940    0.1840    0.5560}

\definecolor{mag}{RGB}{255,0,255}

\begin{document}

\title{When More Data Hurts: Optimizing Data Coverage While Mitigating Diversity Induced Underfitting in an Ultra-Fast Machine-Learned Potential}
\author{Jason B. Gibson}
\email{jasongibson@ufl.edu}
\affiliation{Department of Materials Science and Engineering\char`,~University of Florida\char`,~ Gainesville\char`,~ Florida 32611\char`,~ USA}
\affiliation{Quantum Theory Project\char`,~ University of Florida\char`,~ Gainesville\char`,~ Florida 32611\char`,~ USA}

\author{Tesia D. Janicki}
\affiliation{Sandia National Laboratories\char`,~ Albuquerque\char`,~ NM 87185\char`,~ USA}

\author{Ajinkya C. Hire}
\affiliation{Department of Materials Science and Engineering\char`,~University of Florida\char`,~ Gainesville\char`,~ Florida 32611\char`,~ USA}
\affiliation{Quantum Theory Project\char`,~ University of Florida\char`,~ Gainesville\char`,~ Florida 32611\char`,~ USA}

\author{Chris Bishop}
\affiliation{Sandia National Laboratories\char`,~ Albuquerque\char`,~ NM 87185\char`,~ USA}

\author{J. Matthew D. Lane}
\affiliation{Sandia National Laboratories\char`,~ Albuquerque\char`,~ NM 87185\char`,~ USA}

\author{Richard G. Hennig}
\affiliation{Department of Materials Science and Engineering\char`,~University of Florida\char`,~ Gainesville\char`,~ Florida 32611\char`,~ USA}
\affiliation{Quantum Theory Project\char`,~ University of Florida\char`,~ Gainesville\char`,~ Florida 32611\char`,~ USA}

\date{\today}

\begin{abstract}
Machine-learned interatomic potentials (MLIPs) are becoming an essential tool in materials modeling. However, optimizing the generation of training data used to parameterize the MLIPs remains a significant challenge. This is because MLIPs can fail when encountering local enviroments too different from those present in the training data. The difficulty of determining \textit{a priori} the environments that will be encountered during molecular dynamics (MD) simulation necessitates diverse, high-quality training data. This study investigates how training data diversity affects the performance of MLIPs using the Ultra-Fast Force Field (UF$^3$) to model amorphous silicon nitride. We employ expert and autonomously generated data to create the training data and fit four force-field variants to subsets of the data. Our findings reveal a critical balance in training data diversity: insufficient diversity hinders generalization, while excessive diversity can exceed the MLIP's learning capacity, reducing simulation accuracy. Specifically, we found that the UF$^3$ variant trained on a subset of the training data, in which nitrogen-rich structures were removed, offered vastly better prediction and simulation accuracy than any other variant. By comparing these UF$^3$ variants, we highlight the nuanced requirements for creating accurate MLIPs, emphasizing the importance of application-specific training data to achieve optimal performance in modeling complex material behaviors.

\end{abstract}
\maketitle

Since the inception of machine-learned interatomic potentials (MLIPs)~\cite{org_mlip_1,org_mlip_2}, their use has garnered both excitement and skepticism. The excitement is due to their profound ability to accurately learn and subsequently model a targeted region of the quantum mechanical potential energy landscape (PEL) with the favorable linear scaling of classical empirical potentials~\cite{mlip_over}. The skepticism is due to their lack of an interpretable functional form, which can lead to aberrations when attempting to model regions of the PEL not sampled in the training set used to parameterize the MLIP~\cite{transfer_2017}.

While various forms of MLIPs have been proposed, most leverage either simple representations coupled with complex neural networks~\cite{NN_behler1, NN_roit_1} or complex representations coupled with simple learning models~\cite{snap, qsnap, ace}. The required complexity in either the learning model or representation has hampered the interpretability of the aforementioned MLIPs, which can lead to unidentified holes in the learned PEL~\cite{MLIP_holes}. The simulated aberrations and lack of interpretable functional forms of the MLIPs have led to a community focus on developing high-quality training data~\cite{Eyert2023}. 

Many techniques for \textit{a priori} data generation based on chemical intuition~\cite{mlearn} have been proposed. However, since these methods rely on expert intuition and it can be extremely difficult to know which motifs the MLIP will encounter during simulation, it cannot be guaranteed that the data set has sufficient coverage to enable the modeling of complex material properties or avoid overfitting the MLIP~\cite{small_cell}. This has led to the development of autonomous data generation techniques that maximize the informational entropy of the training data~\cite{EM1, EM2}, and active learning techniques that iteratively refit an MLIP to a growing training set until it can accurately reproduce desired material properties~\cite{sin_mlip_1}. A unifying principle of these methods is to maximize the coverage of the dataset while minimizing the cost of generating the dataset, i.e., the number of \textit{ab inito} calculations.

While this principle may be advantageous to MLIPs that employ neural networks, which abide by the universal approximation theorem~\cite{uat}, it does not account for the finite complexity of MLIPs that employ alternative learning algorithms. This finite complexity can yield an underfit MLIP that lacks the complexity to adequately learn the hypersurface represented by the training data. While it has been argued that learning the high-energy regions of the PEL is unnecessary~\cite{Langer2022}, the issue of underfitting from a dataset that covers too broad a region of the PEL has not been fully explored. We define this critical issue as ``diversity-induced underfitting,'' where the inclusion of a wide variety of training data introduces complexity that exceeds the learning capacity of the MLIP, resulting in suboptimal simulation performance. By recognizing the limits imposed by diversity-induced underfitting, researchers will be better equipped to fine-tune data selection and model complexity, driving improvements in the simulation accuracy of MLIPs across a wide range of applications. In this letter, we aim to elucidate the nuanced requirement of the MLIP training data and identify diversity-induced underfitting by performing an ablation study, which consists of fitting the highly interpretable Ultra-Fast Force-Field (UF$^3$)~\cite{uf3} to subsets of a database consisting of both expert generated and autonomously generated data.

As a test case, we produce an MLIP to model the crystallization of amorphous silicon nitride. Understanding the underlying physical mechanisms that cause the crystallization of amorphous silicon nitride and understanding growth dynamics is of paramount interest, as crystallization can modify the film's electrical, mechanical, and optical properties, which can impact microelectronics fabrication and design. Studying the entire growth process would require simulating systems on the micron scale for hours, which is computationally infeasible. However, studying only the early stages of growth requires simulating systems on the nanoscale for fractions of microseconds, which is accessible with empirical potentials. Unfortunately, existing empirical potentials for silicon nitride~\cite{mg2,tersoff,vashishta} do not offer sufficient transferability for nonstoichiometric interactions~\cite{Tesias_paper}, and previously developed MLIPs~\cite{sin_mlip_1,sin_mlip_2} do not offer sufficient speed.   

To develop an MLIP of sufficient speed and accuracy, we adopt the UF$^3$ framework, which has been shown to accurately model silicon carbide~\cite{MacIsaac2024} and aid in the discovery of hard material in the Si--C--N material system~\cite{Bavdekar2023}. UF$^3$ models the PEL as a sum of effective 2- and 3-body interactions described by
\begin{dmath}\label{eq1}
E =\sum_{i=1}^{N_s}\sum_{j=1}^{N_{l_i}}V_{2}(r_{ij})\\ + \sum_{i=1}^{N_s}\sum_{j=1}^{N_{l_i}}\sum_{k=1}^{N_{l_i}}V_3(r_{ij},r_{ik},r_{jk}),
\end{dmath}
where $N_s$ is the number of atoms in the simulation cell and $N_{l_i}$ is the number of atoms neighboring atom \textit{i}, within a predefined cut-off radius while $r_{ij}$, $r_{ik}$ and $r_{jk}$ are the interatomic distances between respective atoms. The function $V_2$ describes the 2-body interaction, expanded into a linear combination of cubic B-splines, $B_n$,
\begin{dmath}
    V_2(r_{ij}) = \sum_{n=0}^{K}c_nB_n(r_{ij}).
\end{dmath}
The function $V_3$ describes the 3-body interactions by a tensor product of splines,
\begin{dmath}
    V_3(r_{ij},r_{ik},r_{jk}) = \sum_{l=0}^{K_l}\sum_{m=0}^{K_m}\sum_{n=0}^{K_n}c_{lmn}B_l(r_{ij})B_m(r_{ik})B_n(r_{jk}),
\end{dmath}
The number of cubic B-spline basis functions is denoted by $K$, $K_l$, $K_m$, and $K_n$ and $c_n$ and $c_{lmn}$ are the learned weights for 2- and 3-body interaction, respectively. The $\uf$ framework incorporates four regularization terms: one ridge term and one curvature term for each of the 2- and 3-body interactions. The ridge terms control the magnitude of each many-body interaction, while the curvature terms control the smoothness of the learned PEL~\cite{uf3}. This formulation allows for near-quantum accuracy with speeds comparable to typically employed empirical potentials.

\begin{figure}[!h]
    \centering
    \includegraphics[trim={0 0.6cm 0 0}, width=\columnwidth]{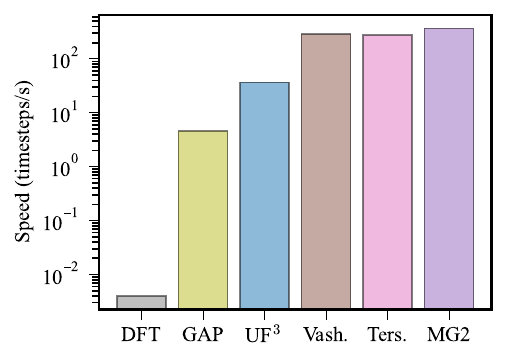}
    \caption{\textbf{Speed comparison of DFT, GAP, empirical potentials and UF$^3$.} The plot shows the speed of DFT, the GAP~\cite{sin_mlip_1} and UF$^3$ MLIP, and the MG2~\cite{mg2}, Vashishta~\cite{vashishta}, and Tersoff~\cite{tersoff} empirical potentials. The speed simulations were ran on 4 CPUs for 100 timesteps with a simulation size of 3024 atoms.}
    \label{speed}
\end{figure}

To demonstrate the speed of UF$^3$, we used 4 Intel Xeon E5-2698 (2.3GHz) CPUs to run an NVE ensemble for 100 timesteps on a 3024 atom supercell of the $\beta$ phase of Si$_3$N$_4$ using the LAMMPS simulation software ~\cite{LAMMPS}. Figure~\ref{speed} compares the speed of UF$^3$ to density functional theory (DFT), the Gaussian approximation potential (GAP)~\cite{GAP}, and other empirical potentials. The DFT and GAP speed is based on the speed reported by Milardovich et al.~\cite{sin_mlip_1}. We observe that while UF$^3$ is slower than the empirical potentials, it is 9,269 times faster than DFT and 8 times faster than GAP. The speed discrepancy between the empirical potentials and UF$^3$ can be attributed to the larger 3-body cutoff radius of UF$^3$. If UF$^3$ adopts the same cutoff radii as the Vashishta potential, the speed of UF$^3$ increases by a factor of 5.

To train $\uf$, we first generated two distinct datasets: the stoichiometric and non-stoichiometric datasets. The stoichiometric dataset was generated following a chemically inspired workflow similar to that of Zou et al.~\cite{mlearn}. We conducted  \textit{ab initio} molecular dynamics (AIMD) simulations at ambient pressure and various temperatures of 300, 1200, 1800, and 3200~K on the primitive cell of the $\alpha$, $\beta$, and $\gamma$ phases of Si$_3$N$_4$. Each simulation was run for 5 ps with a timestep of 1 fs, ensuring structurally diverse data specific to the Si$_3$N$_4$ stochiometry. To ensure sampling of distorted near equilibrium structures, we also included the relaxation trajectories used to compute the elastic constants of the three phases. Lastly, to sample the repulsive regions of atomic interactions, structures sampled from high-pressure relaxation trajectories at up to 200 GPa were also included in the dataset. For the non-stoichiometric datasets, we leveraged a genetic algorithm (GA) structure search using the Genetic Algorithm for Structure and Phase Prediction (GASP) python package~\cite{GASP1, GASP2, GASP3}. Sampling the relaxation trajectories of the GA-produced structures provides an autonomous way to generate structurally and compositionally diverse data while focusing on the low-energy regions of the PEL~\cite{MacIsaac2024,Bavdekar2023}. It is important to note that while we refer to this dataset as non-stoichiometric, it is a super-set of both stoichiometric and non-stoichiometric phases as it still contains stoichiometric phases sampled by the GA. To sample relevant energies and forces, we excluded structures with high forces exceeding 20~eV/\AA\ and energies more than 3~eV/atom above the convex hull. The energy filtration criterion was selected based on the observation that the inclusion of structures with energies above the convex hull greater than 3~eV/atom tended to degrade predictions on low-energy structures, while the force filtration criterion was selected based on the distribution of forces observed while simulating amorphous silicon nitride using the empirical potentials~\cite{mg2,vashishta,tersoff}.

All DFT calculations were run using VASP~\cite{PhysRevB.47.558, PhysRevB.49.14251, KRESSE199615, PhysRevB.54.11169} with the projector augmented wave method~\cite{PhysRevB.50.17953} and the Perdew-Burke-Ernzerhof (PBE) generalized gradient approximation for the exchange-correlation functional~\cite{PhysRevLett.77.3865}. We use a $k$-point density of 45 per \AA$^{-1}$~with the Methfessel-Paxton scheme and smearing of 100~meV for the Brillouin zone integration, and a cutoff energy of 400~eV for the plane-wave basis set. 

To evaluate the effects of diversity-induced underfitting, we fit four variants of UF$^3$. First, we partition the stoichiometric dataset into a test/train split, reserving the 3200~K AIMD trajectory for testing and the remaining data for training and fit one variant to the stoichiometric training set (UF$^3_{\text{stoic.}}$). The 3200~K AIMD was selected as the test dataset to provide a particularly difficult test case as the high-temperature motifs are unlikely present in the training data. Next we partition the non-stoichiometric dataset into an 80/20 train/test split and remove from the test set structures with a nitrogen content greater than the Si$_3$N$_5$ composition. We fit a second variant to the non-stoichiometric training set (UF$^3_{\text{nonstoic.}}$), and a third variant to both the stoichiometric and non-stoichiometric training sets (UF$^3_{\text{total}}$). A final variant was fit to the stoichiometric training set and a subset of the non-stoichiometric training set where structures with a nitrogen content greater than Si$_3$N$_5$ are removed (UF$^3_{\text{final}}$). The UF$^3_{\text{final}}$ variant's hyperparameters were optimized and subsequently used for all other variants.

\begin{figure}[tb]
    \centering
    \includegraphics[trim={0 0.3cm 0 0}, width=\columnwidth]{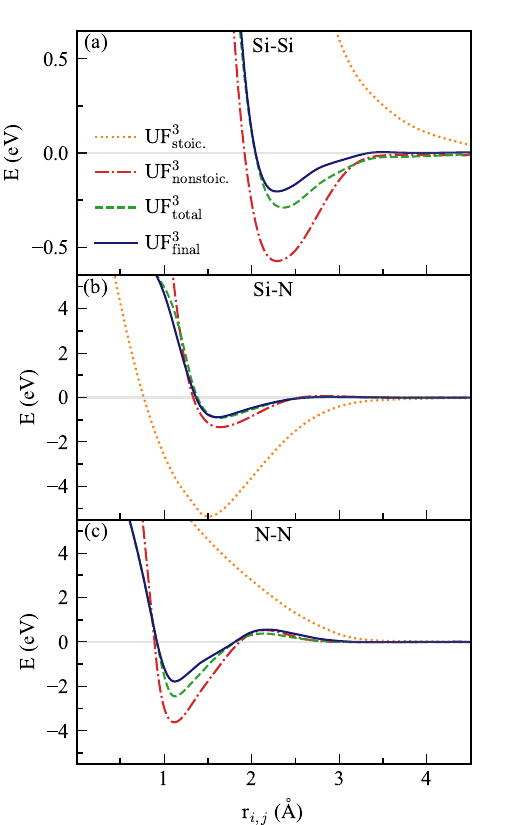}
    \caption{\textbf{Visualization of the 2-body terms for the UF$^3$ variants.} The lines represent the learned two-body interaction for each variant for (a) Si-Si, (b) Si-N, and (c) N-N interactions.}
    \label{uf3}
\end{figure}

A critical advantage of UF$^3$ is the coupling of cubic B-splines to represent atomic interaction with linear regression, which provides easy visualization of the functional form that the MLIP learned. Figure~\ref{uf3} shows the learned 2-body interactions for the four variants. This visualization is an initial check to ensure the MLIP has learned the proper representation of the PEL. For example, UF$^3_{\text{stoic.}}$ has incorrectly learned that Si-Si and N-N interactions are always repulsive. The stoichiometric data set consists primarily of motifs in which silicon is tetrahedrally coordinated with nitrogen, and nitrogen is three-fold coordinated with silicon. Without data on shorter-range N-N or Si-Si interactions, the stoichiometric data set was not sufficiently diverse to enable the MLIP to learn same species attraction.

Because the non-stoichiometric dataset incorporates additional compositions and local geometries, the UF$^3_{\text{nonstoic.}}$, UF$^3_{\text{total}}$, and UF$^3_{\text{final}}$ variants have learned attractive wells for all interactions. Features of these variants, such as the modest barrier around 2~$\AA$ for the N-N interaction that likely prevents N$_2$ sublimation in high-temperature simulations, provide confidence that the MLIP will provide desired physical results. However, further interpretation of these visualizations is rather nuanced without analysis of the force predictions and simulated results. As such, the variants were further validated against the DFT computed forces in the test sets and their ability to reproduce DFT computed elastic constants and the amorphous structure (e.g. density and radial/angular distribution functions).

\begin{figure*}[ht]
    \centering
    \includegraphics[trim={0 1cm 0 0},width=\textwidth]{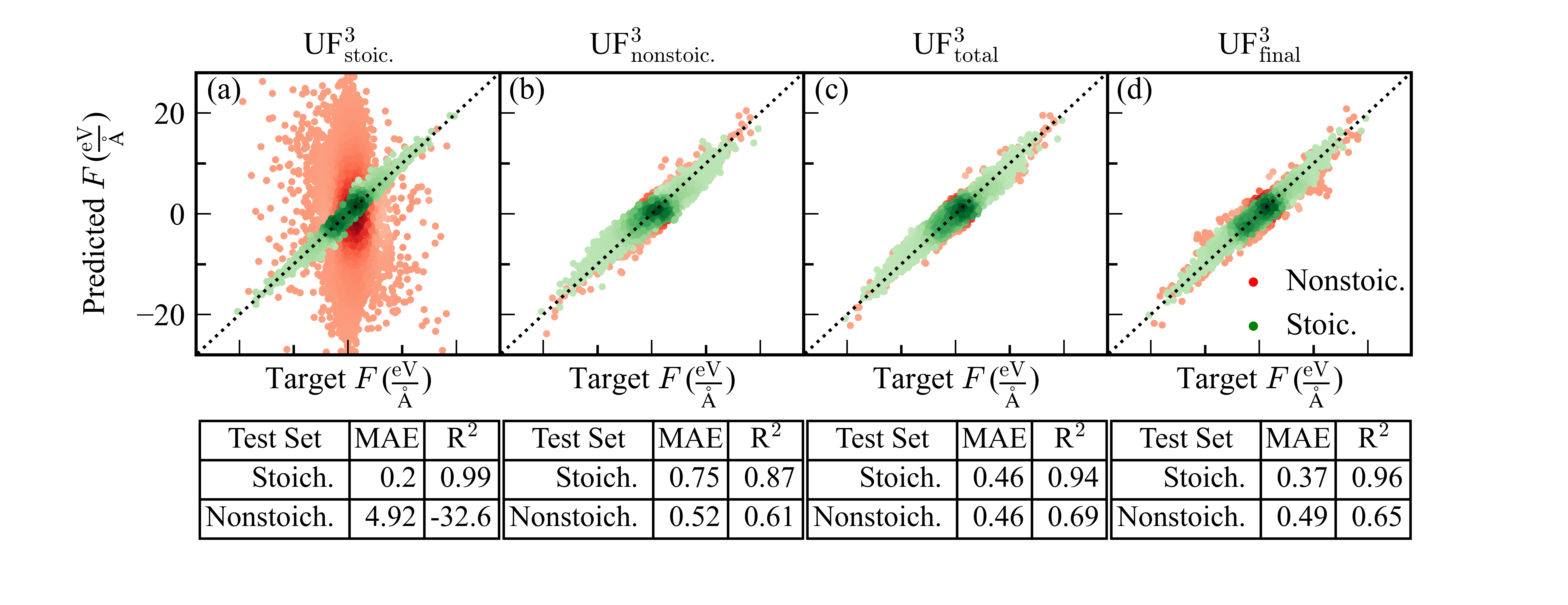}
    \caption{\textbf{Validation force predictions for the $\uf$ variants} The green markers show the predictions for the stoichiometric test set compared to DFT. The red markers show the predictions for the non-stoichiometric test set compared to DFT. The color gradient shows the density of points. The tables show the regression metrics for the variants on both test sets }
    \label{pairity}
\end{figure*}
Figure~\ref{pairity} shows the force predictions and regression metrics compared to DFT for the four UF$^3$ variants. The UF$^3_{\text{nonstoic.}}$, UF$^3_{\text{total}}$, and UF$^3_{\text{final}}$ MLIPs all exhibit reasonable mean absolute errors (MAEs) and $R^2$ values for the force prediction across both test sets, indicating no apparent issues. The UF$^3_{\text{stoic.}}$ performs well for predictions on the stoichiometric test set, obtaining an $R^2$ of 0.99, but is incapable of making even moderately reasonable predictions for the non-stoichiometric test set obtaining an $R^2$ -32.62, as seen in Figure~\ref{pairity}(a). This result shows that UF$^3_{\text{stoic.}}$ will be inadequate for modeling non-stoichiometric interactions. Further, the excellent predictions of UF$^3_{\text{stoic.}}$ on the stoichiometric test set show that test sets drawn from the same narrow distribution as the training set provide a rather biased conclusion of the generalizability of an MLIP. Contrarily, we see that UF$^3_{\text{final}}$ obtained the best overall force predictions as it was trained on a balanced data set.

While UF$^3_{\text{final}}$ did obtain the best overall force predictions, we cannot conclude the variant's quality from force predictions alone. This is because the quality of the MLIP is based on its ability to accurately predict macroscopic observables that describe material properties or reveal detailed physical mechanisms~\cite{fu2023forces}, which requires highly accurate representations of specific regions of the PEL, not just accurate predictions on forces from the distribution it was trained on. As such, we further validate our MLIPs against the simulated properties stated previously.

\begin{figure*}[t]
    \centering
    \includegraphics[trim={0 0.6cm 0 0}, width=\textwidth]{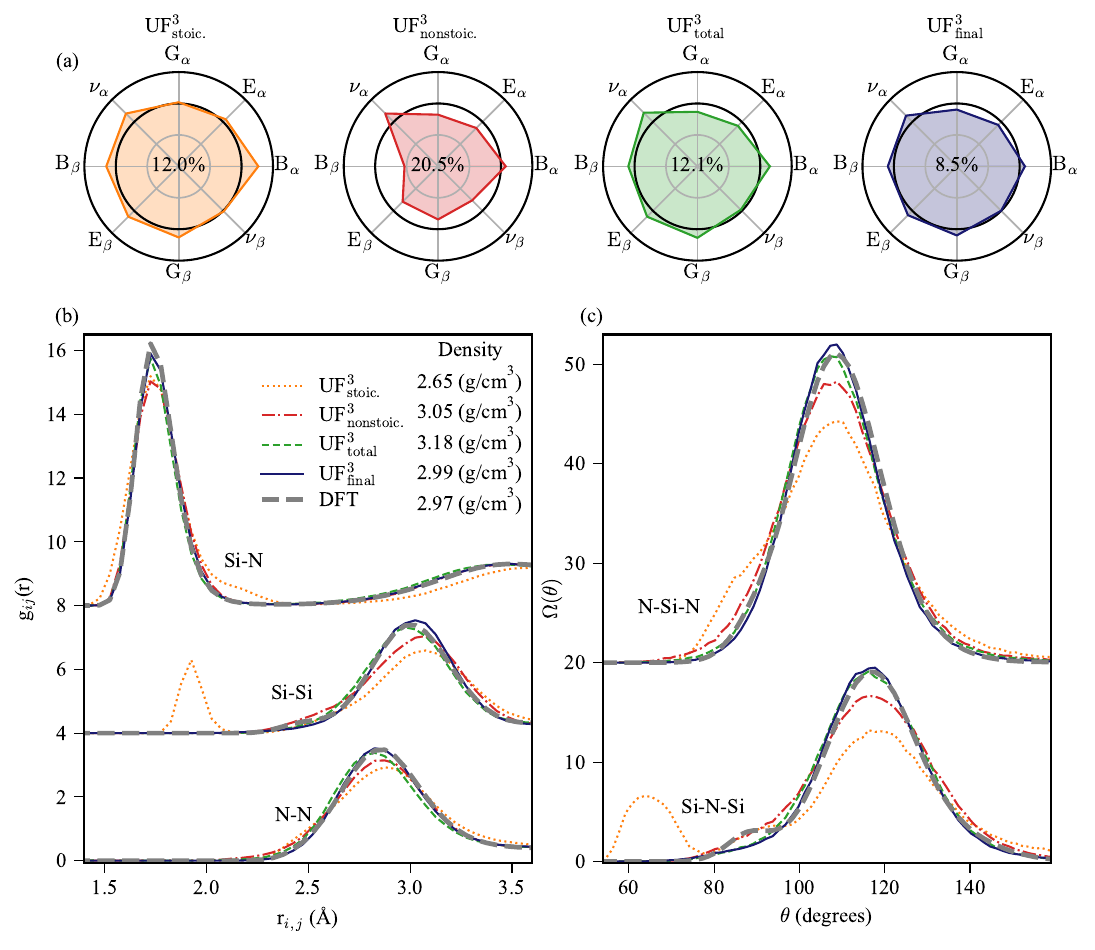}
    \caption{\textbf{Validation simulation results for the $\uf$ variants.} (a) Shows the performance of elastic properties simulated by the UF$^3$ variants compared to DFT. The solid black circle indicates zero error. The inner and outer circles represent 50\% under and over prediction, respectively. The subscript represents the Si$_3$N$_4$ polytype. the number in the spider plots is the mean absolute percent error (MAPE) for all quantities computed compared to DFT. (b,c) Show the structural properties of amorphous Si$_3$N$_4$ at 1700~K for UF$^3$. The initial amorphous structure started from a random configuration of 224 atoms. It was then equilibrated using UF$^3_{\text{final}}$ for 16 ns at 1700 K. This structure was then further equilibrated for an additional 800 ps using either aiMD for the DFT-equilibrated reference or the respective UF$^3$ variant. (b) shows the partial RDF for N-N (bottom), Si-Si (middle), and Si-N (top). The number in the legend is the final density of the amorphous structure. (c) Shows the ADF computed by each variant for N-Si-N (top) and Si-N-Si (bottom) interactions.}
    \label{val}
\end{figure*}

Figure~\ref{val}(a) compares the accuracy of the elastic properties predicted by the four variants for the $\alpha$ and $\beta$ phases. UF$^3_{\text{stoic.}}$, performs well in computing the elastic constants, obtaining a mean absolute percent error (MAPE) of 12.0\%. The performance on the elastic properties is justified as the variant was trained on only stoichiometric motifs, which is what would be encountered in simulating elastic properties.  However, as seen in Fig.~\ref{pairity}, the UF$^3_{\text{stoic.}}$ variant is unable to generalize to nonstoichiometric motifs. This inability to generalize is difficult to identify without an explicit nonstoichiometric test set, as it does not necessarily hinder the MLIP's ability to model stoichiometric interactions. Rather, it underutilizes the flexibility of UF$^3$ and prevents the proper modeling of non-stoichiometric motifs. Improper modeling of non-stoichiometric motifs only becomes apparent when we look at Figs.~\ref{val}(b,c), which shows the amorphous radial and angular distribution function computed by the variant. The UF$^3_{\text{stoic.}}$ MLIP severely under-predicts the density and predicts an overly broad ADF and unphysical peaks at roughly 1.9~\AA\ in the RDF and 65~\textdegree in the ADF. 

Contrarily, although UF$^3_{\text{nonstoic.}}$ shows a modest agreement with DFT for the predictions of the amorphous density, RDF, and ADF, this variant had the highest MAPE for elastic properties, suggesting the model has not learned proper responses to distortions. While the UF$^3_{\text{total}}$ variant had improved performance in computing the elastic properties compared to UF$^3_{\text{nonstoic.}}$, it overestimated amorphous density.

The improved prediction performance for elastic properties but simultaneous degradation in predicting amorphous properties of UF$^3_{\text{total}}$ compared to UF$^3_{\text{nonstoic.}}$ indicates diversity-induced underfitting. If UF$^3_{\text{total}}$ had sufficient complexity to capture the entire training set, it should have improved its predictions of the elastic properties while maintaining reasonable predictions of the amorphous properties, given that the training set simply had additional data specific to the Si$_3$N$_4$ stoichiometry. Nitrogen behavior differs significantly between nitrogen-rich and silicon-rich environments—forming N$_2$ gas in the former and three-fold coordinated nitrogen with silicon in the latter. This diverse behavior within the training set highlights the need for alignment between the training data's diversity and the model's complexity, ensuring balanced and accurate predictions. 

The simulated validation results from UF$^3_{\text{final}}$, which is fit to the stoichiometric data, and filtered non-stoichiometric data, which spans from $\chi_{\mathrm{Si}}$ = 0.375 to $\chi_{\mathrm{Si}}$ = 0.91, provides empirical evidence of the diversity-induced underfitting of UF$^3_{\text{total}}$.
Although UF$^3_{\text{final}}$ was fit to less data, it obtained an MAPE of 8.5\% (Fig.~\ref{val}(a)), which is far better than any of the other variants. Further, the amorphous density, radial distribution function, and angular distribution function shown in Fig~\ref{val}(c) show close agreement with density functional theory. Based on the speed tests, visualizations, force predictions, and simulated validation results, this MLIP is of sufficient speed and accuracy to warrant a detailed investigation of the crystallization of amorphous silicon nitride, which we intend to pursue in a subsequent study. 

Our major conclusion is that the training data requirements of linear MLIPs are far more nuanced than previously thought. Care must be taken when increasing the diversity of the training data. Presenting a region of the potential energy landscape that is too broad for the complexity of an MLIP can lead to diversity-induced underfitting. On the other extreme, a training set consisting only of deformations of crystalline phases will be too easy for an MLIP to learn, yielding erratic simulation behavior when the MLIP is forced to extrapolate to unseen domains. While we recognize it is impossible to determine \textit{a priori} the motifs an MLIP will encounter, our concluding recommendation is that the training data should be sufficiently diverse. Still, paramount effort must be taken to exclude motifs that are clearly irrelevant to the region of the potential energy landscape that the MLIP will explore during simulations to ensure the training data is within the complexity that the MLIP can capture. 

One can accomplish this by following the ablation study outlined in this letter. If removing specific subsets of the training yields more accurate simulated results, the original dataset likely spans too broad a region of the PEL for the MLIP to capture correctly and the MLIP suffers from diversity-induced underfitting. Such is the case when comparing UF$^3_{\text{total}}$ to UF$^3_{\text{final}}$. Alternatively, if any of the simulated validation results portray aberrations, regardless of whether other validation results show close agreement with DFT, it is likely the training is not sufficiently diverse to produce an MLIP that can generalize for the desired application. Such is the case for UF$^3_{\text{stoic.}}$ where the MLIP provided reasonable predictions of the elastic properties but predicted unphysical peaks in the amorphous radial distribution function. While we focus our investigation on the UF$^3$ machine learned interatomic potential, we expect the findings and recommendations of this letter to generalize to any MLIP of finite complexity.

The authors acknowledge the helpful insights and discussion with the entire UF$^3$ development team. Part of this research was performed while J.B.G., A.C.H., and R.G.H. were visiting the Institute for Pure and Applied Mathematics (IPAM), which is supported by the National Science Foundation (Grant No. DMS-1925919). This work was supported by the Laboratory Directed Research and Development program at Sandia National Laboratories, a multimission laboratory managed and operated by National Technology and Engineering Solutions of Sandia, LLC, a wholly owned subsidiary of Honeywell International Inc., for the U.S. Department of Energy's National Nuclear Security Administration under contract DE-NA0003525. This paper describes objective technical results and analysis. Any subjective views or opinions that might be expressed in the paper do not necessarily represent the views of the U.S. Department of Energy or the United States Government.

\bibliography{references}

\end{document}


\title{Supplemental material: Accelerating superconductor discovery through modern deep learning  of the electron-phonon spectral function}

\author{Jason Gibson}
\email{jasongibson@ufl.edu}
\affiliation{Department of Materials Science and Engineering, University of Florida, Gainesville, Florida 32611, USA}
\affiliation{Quantum Theory Project, University of Florida, Gainesville, Florida 32611, USA}

\author{Tesia Janicki}
\affiliation{Sandia Nation Laboratories, Albuquerque, NM 87185, USA}

\author{Ajinkya Hire}
\affiliation{Department of Materials Science and Engineering, University of Florida, Gainesville, Florida 32611, USA}
\affiliation{Quantum Theory Project, University of Florida, Gainesville, Florida 32611, USA}

\author{Chris Bishop}
\affiliation{Sandia Nation Laboratories, Albuquerque, NM 87185, USA}

\author{J. Matthew D. Lane}
\affiliation{Sandia Nation Laboratories, Albuquerque, NM 87185, USA}

\author{Richard G. Hennig}
\affiliation{Department of Materials Science and Engineering, University of Florida, Gainesville, Florida 32611, USA}
\affiliation{Quantum Theory Project, University of Florida, Gainesville, Florida 32611, USA}
\affiliation{Department of Physics, University of Florida, Gainesville, Florida 32611, USA}

\maketitle


\section{Supplemental Note 1: comparing out a2f calculations}

\begin{figure}[h]
    \centering
    \includegraphics[width=\columnwidth]{figures/a2F_comparison.pdf}
    \caption{Calculated $\atf$ for Nb, MgB$_2$, and PdH as compared to the literature}
    \label{a2F_comp_w_lit}
\end{figure}

Here we compare our $\atf$ with those from literature (digiized using WebPlotDigitizer~\cite{Rohatgi2022}) and $\atf$ calculated by Cerqueria \emph{et al.}~\cite{Cerqueira2023} for three materials- Nb, MgB$_2$, and PdH. The $\atf$ from literature use a much dense k and q-grid as compared to us and Cerqueria \emph{et al.} For simple elemental system like Nb the converged calculations from \emph{Wang et al.}~\cite{Wang2020} agree fairly well with our $\atf$, but for MgB$_2$ and PdH the disagreement is large.  
Supplementary Table~\ref{supp_table1} shows the electron-phonon coupling constant, shape factors ($\omega_{\text{log}}$ and $\omega_2$) and Allen-Dynes $T_c$ for the three materials. The AD $Tc$ for Nb, MgB$_2$ and PdH calculated using $\alpha^2F(\omega)$ presented in this work agree well with the $T_c$s computed from the $\atf$ published in the literature.

\begin{table}
\caption{Calculated $\lambda$, $\omega_{\text{log}}$, $\omega_2$ and AD T$_c$. $\mu^*=0.1$ was used in the Allen-Dynes equation}
\begin{tabular}{c|c|c|c|c|c}
\hline
Material              & Source           & $\lambda$ & $\omega_{\text{log}}$ & $\omega_2$ & AD T$_c$ \\ \hline
\multirow{3}{*}{Nb}   & This work        & 1.23   & 168.9      & 195.2  & 16.9  \\
                      & Cerqueria et al. & 1.11   & 179.1      & 206.3  & 15.6  \\
                      & Wang et al.\cite{Wang2020}      & 1.51   & 129.8      & 173.6  & 16.9  \\ \hline
\multirow{3}{*}{MgB2} & This work        & 0.62   & 759.2      & 816.3  & 20    \\
                      & Cerqueria et al. & 0.52   & 780.5      & 843.0  & 10.9  \\
                      & Poncé et al.\cite{Ponc2016}     & 0.75   & 641.2      & 708.2  & 27.32 \\ \hline
\multirow{3}{*}{PdH}  & This work        & 1.58   & 300.4      & 391.8  & 41.1  \\
                      & Cerqueria et al. & 1.52   & 313.5      & 398.6  & 41.1  \\
                      & Errea et al.\cite{Errea2013}     & 1.55   & 298.6      & 399.5  & 41.6  \\ \hline
\end{tabular}
\label{supp_table1}
\end{table}


\newpage
\section{Supplemental Note 2: Training and Testing split}

\newpage
\section{Supplemental Note 3: Comparing derived properties for smoothed and raw a2f}

\newpage
\section{Supplemental Note 5: Applying CSO model}

\bibliography{references}